\documentclass[a4paper,12pt, oneside]{article}
\usepackage[dvipdfm]{graphicx}% Include figure files
\usepackage{dcolumn}% Align table columns on decimal point
\usepackage{bm}% bold math
\begin{document}
\baselineskip 24pt
%\begin{spacing}{2}

\begin{center}
{\Large 
Ferromagnetism in ZnO co-doped with Mn and N studied by 
soft x-ray magnetic circular dichroism}
\vspace{1cm}

\rm{T.~Kataoka, Y.~Yamazaki, V.~R.~Singh and Y.~Sakamoto}\\
\it{
Department of Physics and Department of Complexity Science and Engineering, 
University of Tokyo, Bunkyo-ku, Tokyo 113-0033, Japan
}%
\vspace{0.5cm}

\rm{A. Fujimori}\\
\it{
Department of Physics and Department of Complexity Science and Engineering, University of Tokyo, Bunkyo-ku, Tokyo 113-0033, Japan;\\ 
Synchrotron Radiation Research Unit, Japan Atomic Energy Agency, Sayo-gun, Hyogo 679-5148, Japan
}%
\vspace{0.5cm}
%%%%%%%%%%%%%%%%%%%%%%%%%%%%%%%%%%%%%%%%%%%%%%%%%%%%%%%%%%%%%%%%%%%%%%%%%

\rm{Y. Takeda, T. Ohkochi, S.-I. Fujimori, T. Okane and Y. Saitoh}\\
\it{
Synchrotron Radiation Research Unit, Japan Atomic Energy Agency, Sayo-gun, Hyogo 679-5148, Japan
}%
\vspace{0.5cm}
%%%%%%%%%%%%%%%%%%%%%%%%%%%%%%%%%%%%%%%%%%%%%%%%%%%%%%%%%%%%%%%%%%%%%%%%%%%%%

\rm{H. Yamagami}\\
\it{
Synchrotron Radiation Research Unit, Japan Atomic Energy Agency, Sayo-gun, Hyogo 679-5148, Japan;\\
Department of Physics, Faculty of Science, Kyoto Sangyo University, Kyoto 603-8555, Japan
}%
\vspace{0.5cm}
%%%%%%%%%%%%%%%%%%%%%%%%%%%%%%%%%%%%%%%%%%%%%%%%%%%%%%%%%%%%%%%%%%%%%%%%%%%%%%%

\rm{A.~Tanaka}\\
\it{Department of Quantum Matter, ADSM, Hiroshima University, Higashi-Hiroshima 739-8530, Japan
}%
\vspace{0.5cm}

\rm{M.~Kapilashrami, L.~Belova and K. V.~Rao}\\
\it{Department of Materials Science-Tmfy-MSE, Royal Institute of Technology, SE 10044 Stockholm, Sweden
}%
\vspace{0.5cm}

\end{center}

\date{\today}

\newpage
\begin{center}
\section*{Abstract}
\end{center}
We have investigated the electronic structure of ZnO:Mn and ZnO:Mn,N thin films using x-ray magnetic circular
dichroism (XMCD) and resonance-photoemission spectroscopy. 
From the Mn 2$p$$\rightarrow$3$d$ XMCD results, 
it is shown that, while XMCD signals only due to paramagnetic Mn$^{2+}$ ions were observed in ZnO:Mn, 
nonmagnetic, paramagnetic and ferromagnetic Mn$^{2+}$ ions coexist in ZnO:Mn,N. 
XMCD signals of ZnO:Mn,N revealed that the localized Mn$^{2+}$ ground state and Mn$^{2+}$ state 
hybridized with ligand hole coexisted, implying $p$-$d$ exchange coupling. 
In the valence-band spectra, spectral weight near the Fermi level was suppressed, 
suggesting that interaction between magnetic moments in ZnO:Mn,N has localized nature.

\newpage

%%\section*{I. Introduction}

In the field of spintronics, it is essential to develop  diluted magnetic semiconductors (DMSs) with ferromagnetism at room temperature. 
Theoretical study \cite{Dietl} based on Zener's $p$-$d$ exchange model has 
predicted that wide-gap semiconductors such as Mn-doped ZnO (ZnO:Mn) are 
promising candidates for room temperature ferromagnetic DMSs. 
However, experimental results obtained so far have not been mutually consistent, i.e., 
some studies \cite{Sharma, Singhala, Jayakumar} reported the observation of intrinsic ferromagnetism 
whereas others \cite{Droubay, Garcia, Lawes} reported the absence of ferromagnetism or extrinsic ferromagnetism in these materials. 
Recently, there has been works \cite{Coey, Kolesnik} suggesting that structurally perfect ZnO:Mn DMSs do not exhibit ferromagnetic order. 
According to first-principles calculations \cite{Sato}, 
if carriers are not doped into ZnO:Mn, antiferromagnetic superexchange interaction predominates so that 
the ground state is an insulating antiferromagnetic spin glass 
while ferromagnetism can be stabilized only by hole doping, e.g., through N substitution for O. 
These reports imply that hole doping is necessary for ferromagnetism in ZnO:Mn. 
Indeed, ferromagnetism has been reported for N-doped ZnO:Mn (ZnO:Mn,N) \cite{Kittilstved, Zhao}. 
However, ferromagnetism has also been observed for $n$-type ZnO:Mn \cite{Norton}. 
It is thus debatable whether ferromagnetism occurs only in a $p$-type ZnO:Mn or not. 
In order to clarify the above issue, x-ray magnetic circular dichroism (XMCD) is an ideal tool 
because XMCD is an element specific probe and the line shapes of XMCD spectra are fingerprints of magnetically active components. 
In particular, from the magnetic field dependence of XMCD signals, 
one can judge whether the XMCD signals come from ferromagnetism or not \cite{Tietze}.  
In this paper, we show the magnetic field dependence of Mn 2$p$$\rightarrow$3$d$ XMCD and 
resonance-photoemission spectroscpy spectra of  the ZnO:Mn,N sample and 
discuss the magnetic states of the doped Mn ions and their electronic structures.

%%\section*{II. Experimental}

400-nm-thick thin films of ZnO:Mn and its N-doped one (Mn=1-2\%) 
were grown on Si substrates by reactive sputtering using 99.9\% pure 
Zn target using a radio frequency (RF)/direct current (DC) sputtering system. 
The substrate temperature during deposition was maintained at 350 $^{\circ}$C. 
The ZnO:Mn,N film was prepared under N pressure of $P_{\rm N_2}$ = 1.5 mbar. 
The films were deposited from separate metallic targets of Zn and Mn 
in our sputtering system which can accommodate up to three targets simultaneously. 
These metal targets were independently characterized prior to the deposition, 
and found no evidence for any contamination. 
From low field ($<$ 50 Oe), zero field cooled and field cooled measurements, 
we found no evidence for such phases and ruled out possible existence of Mn oxide clusters as well. 
Moreover, we rule out impurity phases in the films by using the X-ray photoelectron spectroscopy (XPS) and 
X-ray diffraction (XRD) measurements. 
X-ray absorption spectroscopy (XAS) and XMCD measurements were performed at the undulator beam line BL-23SU of SPring-8 and 
spectra were recorded in the total-electron-yield (TEY) mode (probing depth $\sim$ 5 nm). 
The degree of circular polarization was higher than $\sim$ 95\%. 
The monochromator resolution was $E$$/$$\Delta$$E$ $>$ 10000. 
Magnetic fields $H$ up to 8 T were applied perpendicular to the sample surface. 
RPES measurements in the Mn 2$p$-3$d$ core-excitation regions were performed at BL-23SU of SPring-8. 
All binding energies (E$_B$) were referenced to 
the Fermi level (E$_F$) of the sample holder which was in electrical contact with the sample. 
The total energy resolution of the RPES measurements was $\sim$170 meV.

%%\section*{III. Results and discussion}

Figure 1(a) shows comparison of the Mn 2$p$$\rightarrow$3$d$ XAS spectra of the ZnO:Mn and ZnO:Mn,N samples at $H$=0 T 
with cluster-model calculations \cite{Tanaka} for the high-spin Mn$^{2+}$, Mn$^{3+}$, and Mn$^{4+}$ states in the $T_d$-symmetry crystal field. 
From the line-shape analysis, we suggest that the doped Mn ions in ZnO:Mn and ZnO:Mn,N are mostly in the high-spin Mn$^{2+}$ states. 
Next, we discuss the magnetically active component of ZnO:Mn and ZnO:Mn,N. 
Figures 1(b) and (c) show the Mn 2$p$-3$d$ XAS spectra, taken at $H$=8 T, using circular polarized x-rays 
and their difference spectrum, i.e., XMCD spectra. 
Here, $\mu$$^+$ and $\mu$$^-$ refer to absorption spectra for photon helicity parallel and 
antiparallel to the Mn 3$d$ spin, respectively. 
The XMCD peak positions of ZnO:Mn,N are the same as the XAS peak positions, indicating that 
the XMCD signal is mainly due to the Mn$^{2+}$ states. 
Considering this and that the XMCD line shape of the ZnO:Mn,N  sample is similar to that of the ZnO:Mn sample, 
we conclude that the magnetically active components both in ZnO:Mn and ZnO:Mn,N mainly come from the Mn$^{2+}$ states. 

Figure 2(a) shows the Mn 2$p$$\rightarrow$3$d$ XMCD spectra of the ZnO:Mn sample 
measured at various magnetic fields at $T$=10 K. 
It is shown that the peak positions of the XMCD spectra are independent of magnetic field. 
The intensities of XMCD signals are negligibly small at $H$=0.1 T and the XMCD signals increase with magnetic field. 
The XMCD intensities as a function of magnetic field are plotted in Fig. 2(b). 
This figure indicates that the paramagnetic Mn$^{2+}$ state exists in this sample and 
the ferromagnetic component is negligibly small because XMCD signals do not 
persist at $H$=0 as a remanence magnetization.
%%In addition to this, the magnetic field dependence of the XMCD intensity shows a convex behavior, 
%%suggesting the presence of an antiferromagnetic component.
%However, we note that the XMCD measurement taken in the TEY-mode is surface sensitive and 
%the magnetically dead layer may exist in the surface region.

Figure 3(a) shows the Mn 2$p$$\rightarrow$3$d$ XMCD spectra of the ZnO:Mn,N sample 
measured at various magnetic fields at $T$=10 K. 
We find that the paramagnetic Mn$^{2+}$ state exists in this sample like in the ZnO:Mn sample, 
because the XMCD signals increase with magnetic field, 
and the peak positions of the XMCD spectra are independent of magnetic field. 
However, one can observe a small but clear XMCD signal of ZnO:Mn,N at $H$=0.1 T as shown in Fig. 3(b). 
Considering that the peak positions of the XMCD spectra are independent of magnetic field, 
the figure suggests the presence of ferromagnetic component due to the Mn$^{2+}$ states. 
To confirm the ferromagnetism, the Mn 2$p$$\rightarrow$3$d$ XMCD spectra 
measured at various magnetic fields at $T$=300 K are shown in Fig. 3(c). 
According to a Currie-Weiss law, it is expected that the paramagnetic component is reduced at room temperature and one can clearly see the ferromagnetic component. 
In Fig. 3(d), the finite XMCD signals persist at $H$ = 0 as a remanence magnetization, suggesting 
the presence of a small ferromagnetic component. 
From the magnetic field dependence of XMCD intensities, 
we have estimated the relative concentrations of the nonmagnetic, paramagnetic and ferromagnetic Mn ions to be $\sim$ 66, 33, and 1\%, respectively. 
This indicates that most of Mn ions are in an effectively non-magnetic state.

Now, we discuss the origin of the ferromagnetic component of ZnO:Mn,N. 
Considering that the ferromagnetic component of XMCD signal is due to Mn$^{2+}$ ions, and MnO (Mn$^{2+}$) is not ferromagnetic material \cite{Nesbet}, 
ferromagnetism due to secondary phases seems unlikely. 
Furthermore, as shown in Fig. 3(c), one can notice features denoted $\alpha$, which are located at 0.5 eV lower photon energy with respect to the main peak. 
Considering the cluster-model calculation by Edmonds ${et}$ ${al}$. \cite{Edmonds} and an XMCD study \cite{Takeda} of GaAs:Mn, 
we suggest that the feature denoted $\alpha$ comes from a Mn$^{2+}$ state hybridized with ligand-hole states, 
whereas the peak around $h$$\nu$$\sim$641.5 eV originates from the localized Mn$^{2+}$ ground state. 
In the case of $p$-type doping, the Mn-Mn exchange coupling may be mediated by the presence of holes in the valence band 
which hybridize with the 3$d$ states of Mn. 
The itinerant holes thereby retain their $d$-like character. 
Therefore, hybridization between Mn 3$d$ and ligand hole stabilizes the ferromagnetic phase for Mn doping, i.e., $p$-$d$ exchange coupling. 
Assuming that hole is doped into the ZnO:Mn,N film through N substitution for O, 
we suggest that the observed ferromagnetism is due to $p$-$d$ exchange coupling. 
If the doped holes are itinerant, it is expected that partial density of states (PDOS) due to Mn 3$d$ states have a finite value at Fermi level \cite{Iusan}. 

Therefore, we have performed the valence-photoemission study of ZnO:Mn,N as shown in Fig. 4(a). 
All the spectra were taken using photon energies in the Mn 2$p$$\rightarrow$3$d$ core-excitation region. 
Labels A to G in the Mn 2$p$$\rightarrow$3$d$ XAS spectrum [Fig. 4(b)] indicate photon energies at which the spectra in Fig. 4(a) were taken. 
By subtracting the off-resonance spectrum (A) from on-resonance one (D), 
we have deduced the Mn 3$d$ PDOS as shown in Fig. 4(c). 
Here, we also compare the experimental Mn 3$d$ PDOS with the theoretical spectrum calculated using the cluster model. 
The calculated spectrum has well reproduced with the experimental one, indicating that we have succeeded in deducing the Mn 3$d$ PDOS. 
For the Mn 3$d$ PDOS, no photoemission intensity was observed at the Fermi level. 
It is likely that the  $p$-$d$ exchange interaction in ZnO:Mn,N has the localized nature, 
reflecting a small amount of the doped N ions.

%\section*{IV. Summary}
%Conclusion

In conclusion, while XMCD signals only due to the paramagnetic component of the Mn$^{2+}$ ions were observed for ZnO:Mn, 
XMCD signals due to the paramagnetism and ferromagnetism of the Mn$^{2+}$ ions were observed for ZnO:Mn,N. 
The estimated relative concentrations of the nonmagnetic, paramagnetic and ferromagnetic Mn ions were $\sim$ 66, 33, and 1\%, respectively, 
in the ZnO:Mn,N sample. 
Also, XMCD signal of ZnO:Mn,N at 300 K revealed that the localized Mn$^{2+}$ ground state and the Mn$^{2+}$ state 
hybridized with ligand hole coexisted, implying the importance of $p$-$d$ exchange coupling for the ferromagnetism of ZnO:Mn,N. 
From the RPES results, we suggest that the $p$-$d$ exchange interaction in ZnO:Mn,N has localized nature, 
reflecting the small amount of the doped N ions.

%\newpage
%%\section*{Acknowledgement}
%Acknowledgements
The experiment at SPring-8 was performed under the approval of the Japan 
Synchrotron Radiation Research Institute (JASRI) (proposal no. 2008A3825). 
The work in Japan was supported by a Grant-in-Aid for Scientific Research (S22224005) from JSPS, Japan. 
The work in Stockholm was funded by the Swedish Agencies VINNOVA and SSF.

\section*{Figures}
\end{center}

\begin{figure}[htbp]
 \begin{center}
  \includegraphics[width=\linewidth]{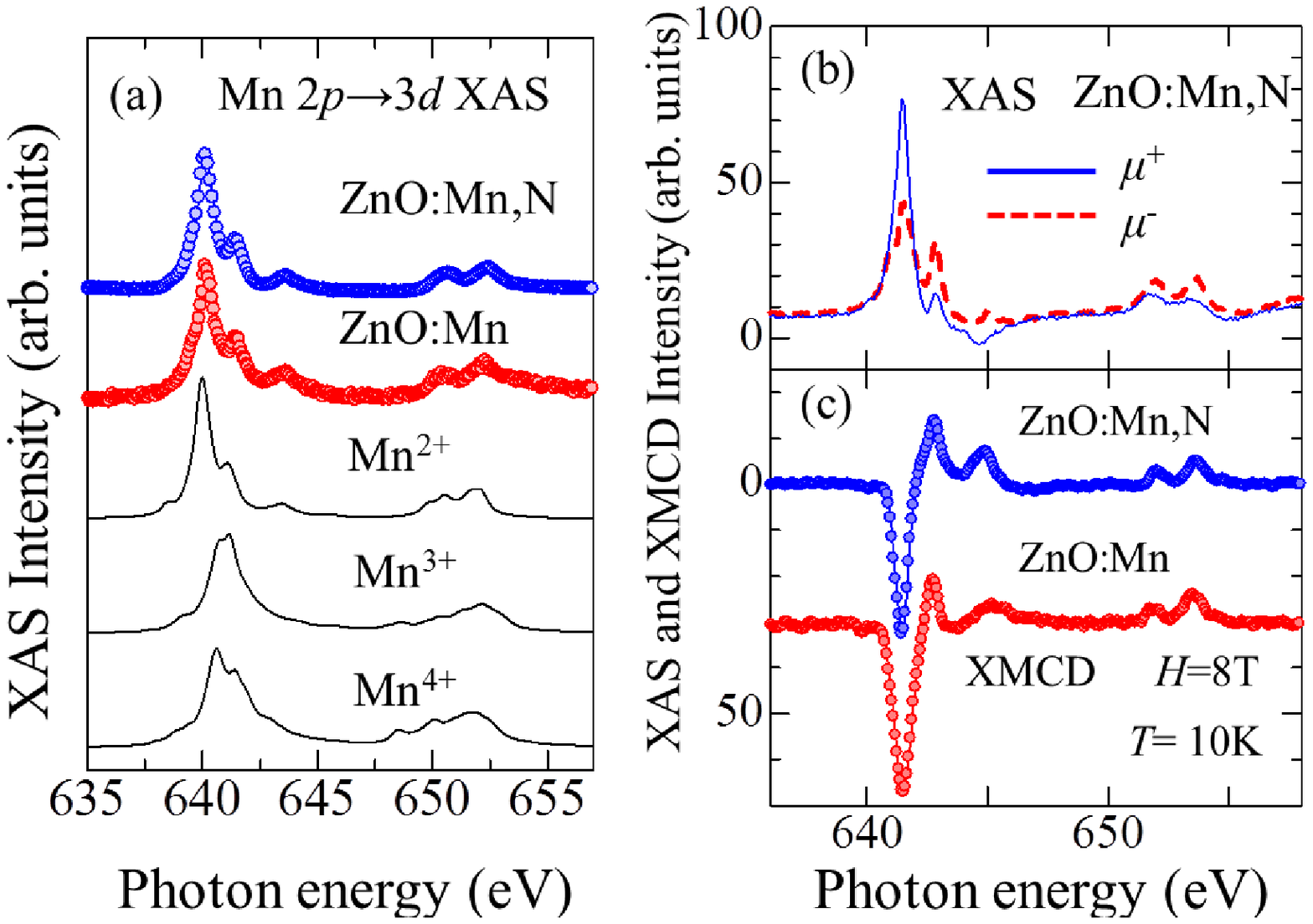}
 \end{center}
 \caption{(Color online) 
Mn 2$p$$\rightarrow$3$d$ XAS and XMCD spectra of ZnO:Mn and ZnO:Mn,N. 
(a) Experimental XAS spectra and theoretical spectra computed using the cluster model. 
(b) XAS spectra of ZnO:Mn,N, taken at $H$=8 T. 
(c) XMCD spectra of ZnO:Mn and ZnO:Mn,N taken at $H$=8 T.}
\end{figure}

\begin{figure}[htbp]
 \begin{center}
  \includegraphics[width=\linewidth]{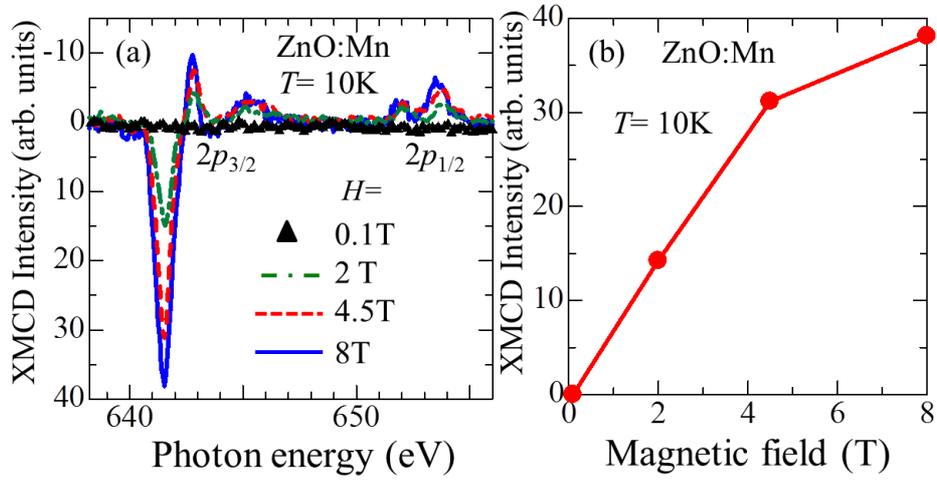}
 \end{center}
 \caption{(Color online) 
Magnetic field dependence of Mn 2$p$$\rightarrow$3$d$ XMCD spectra for ZnO:Mn. 
(a) XMCD at various $H$ taken at $T$=10 K. 
(b) Intensity of Mn 2$p$$\rightarrow$3$d$ XMCD at $T$=10 K as a function of $H$.}
\end{figure}

\begin{figure}[htbp]
 \begin{center}
  \includegraphics[width=\linewidth]{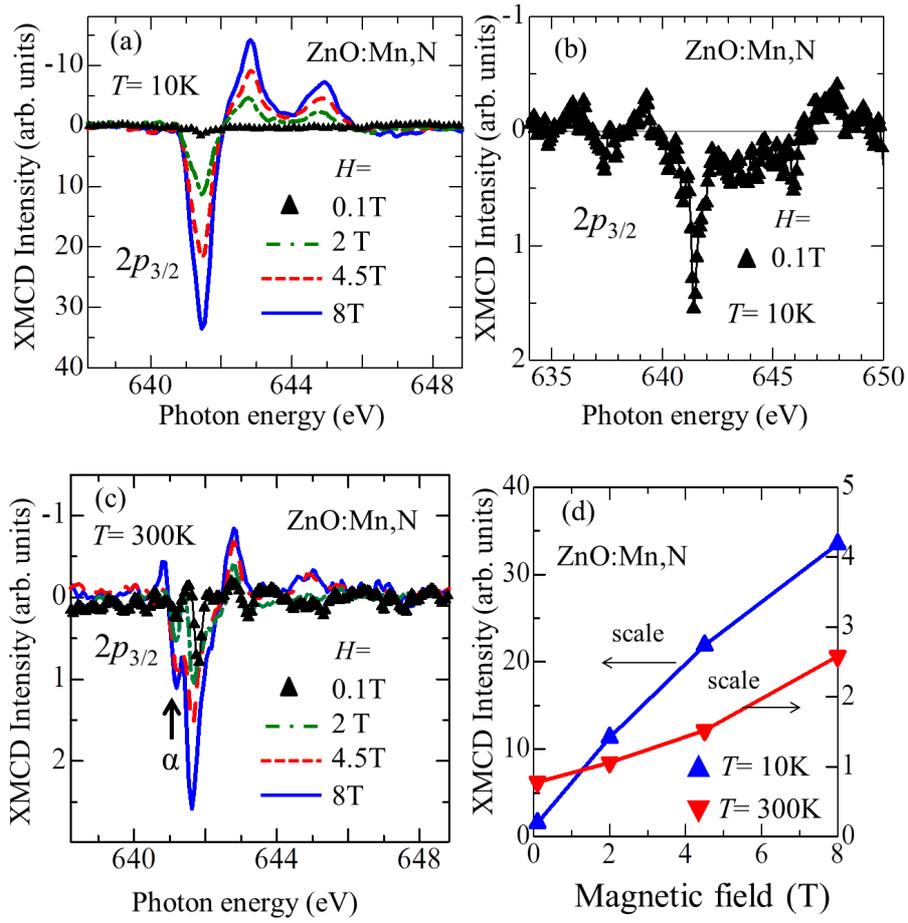}
 \end{center}
 \caption{(Color online) 
Magnetic field dependence of Mn 2$p$$\rightarrow$3$d$ XMCD for ZnO:Mn,N. 
XMCD at various $H$ taken at $T$=10 (a) and 300 K (c). 
(b) Magnified view of XMCD spectrum of ZnO:Mn,N at $H$=0.1T and $T$= 10 K.
(d) Intensity of XMCD spectra of ZnO:Mn,N at $T$=10 and 300 K as functions of $H$.}
\end{figure}

\begin{figure}[htbp]
 \begin{center}
  \includegraphics[width=\linewidth]{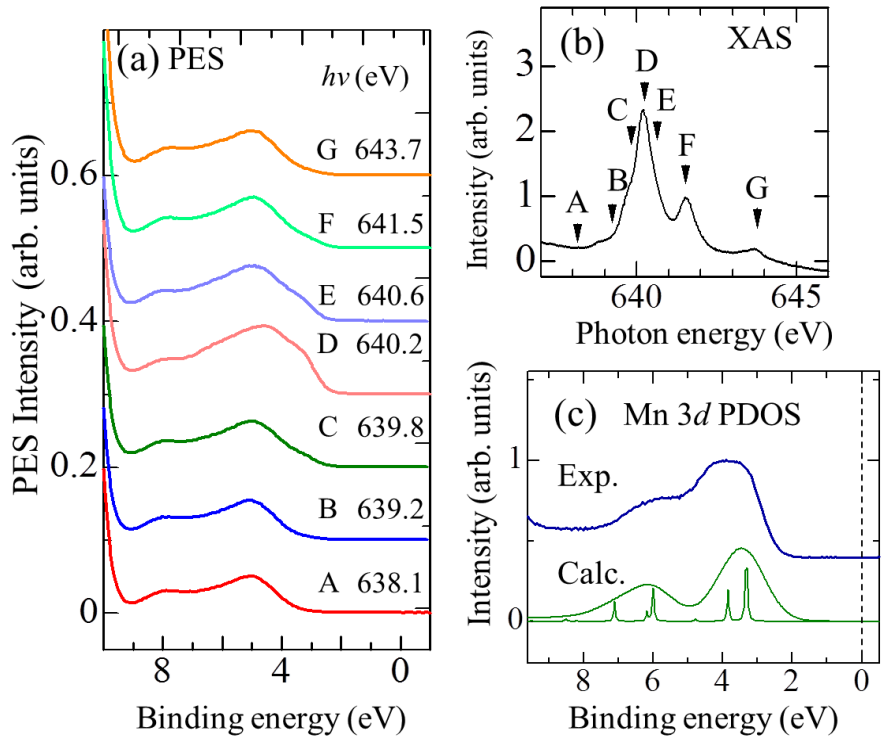}
 \end{center}
 \caption{(Color online) 
Valence-band PES spectra (a) of ZnO:Mn,N taken with various photon energies.  (b) Mn 2$p$$\rightarrow$3$d$ XAS spectrum. 
(c) Experimental Mn 3$d$ PDOS and theoretical one computed using the cluster model.
The electronic structure parameters for the calculation were taken from the literature \cite{Mizokawa} on the photoemission study of ZnO:Mn.}
\end{figure}

\end{document}